# A novel feed rate scheduling method based on Sigmoid function with chord error and kinematics constraints

Hexiong Li[1,2], Xin Jiang[1,2,3,*], Guanying Huo[1,2], Cheng Su[1,2], Bolun Wang[1,2], Yifei Hu[1,2], Zhiming Zheng[1,2,3]

[1]Key Laboratory of Mathematics, Informatics and Behavioral Semantics (LMIB), School of Mathematics Science, Beihang University, Beijing, China

[2]Peng Cheng Laboratory, Shenzhen, Guangdong, China

[3]Beijing Advanced Innovation Center for Big Data and Brain Computing (BDBC), Beihang University, Beijing, China

Corresponding author: Xin Jiang (e-mail: jiangxin@buaa.edu.cn ).

**Abstract:**

In high speed CNC (Compute Numerical Control) machining, the feed rate scheduling has played an important role to ensure machining quality and machining efficiency. In this paper, a novel feed rate scheduling method is proposed for generating smooth feed rate profile conveniently with the consideration of both geometric error and kinematic error. First, a relationship between feed rate value and chord error is applied to determine the feed rate curve. Then, breaking points, which can split whole curve into several blocks, can be found out using proposed two step screening method. For every block, a feed rate profile based on the Sigmoid function is generated. With the consideration of kinematic limitation and machining efficiency, a time-optimal feed rate adjustment algorithm is proposed to further adjust feed rate value at breaking points. After planning feed rate profile for each block, all blocks feed rate profile will be connected smoothly. The resulting feed rate profile is more concise compared with the polynomial profile and more efficient compared with the trigonometric profile. Finally, simulations with two free-form NURBS curves are conducted and comparison with the sine-curve method are carried out to verify the feasibility and applicability of the proposed method.

feed rate adjusting

## Introduction

Feed rate scheduling, which aims to generate smooth feed rate profile of tool motion under geometric and kinematic limitations, is one of the most important process in the parametric interpolation, and has important impact on machining process especially in the term of machining efficiency and quality. It is obviously that greater feed rate value leads to shorter machining time. However, on the other hand, the feed rate, which exceed the limitation, will cause significant geometric deviation [1,2]. Specifically, excessive feed rate value can cause larger contour error [1] and chord error [2]. Therefore, feed rate scheduling needs to determine whether the speed should be increased to improve efficiency or decreased to satisfy kinematic constraints [3]. Meanwhile, feed rate scheduling method also should be succinct and convenient to implement.

Nowadays, a significant number of feed rate scheduling methods have been proposed, and the existing feed rate scheduling method can be roughly classified as two approaches: time-optimal approach and ACC/DEC approach.

For time-optimal approach, a minimal-time control problem, which is confined with given error, is present and the shape of feed rate profile can be determined by solving differential equations. There are two approaches to deal with differential equations: analytic approach and numeric approach. Timar and Farouki[4] obtained time-optimal feed rate functions under constant or speed dependent acceleration limits by solving two differential equations, which have closed-form solutions. In the research of Zhang et al. [5], they reduced the chord error bound to a centripetal acceleration bound which leads to a velocity limit curve, called the chord error velocity limit curve. According to "bang-bang" control principle [6], the velocity limit curve was the time-optimal curve. Although analytic approach can achieve time-optimal feed rate profile, it is difficult to be applied in multiple constraints and high-order constraints time-optimal problems. At this time, numerical approach is used. Dong and Stori [7] proposed an optimal formulate with two types of constraints which were the equations explicitly including parametric acceleration and functions only of the path geometry and parametric velocity. The formulate used bidirectional scan algorithm to solve time-optimal feed rate, which satisfied acceleration and contour error constraints. Many researchers have proposed various numerical time-optimal method with kinds of constraints. In the article of Sun et, al. [8], chord error and kinematic constraints including feed rate, acceleration and jerk constraints were involved, and the feed rate value at every point was adjusted iteratively by multiplying constant coefficient. Bharathi and Dong [9] proposed heuristic smooth feed rate optimization algorithm with high-order constraints. Lu et, al. [10] used an efficient numerical method based on Pontryagin maximum principle to solve time-optimal problems with tool-tip kinematic constraints. Ye et al. [11] re-parameterized tool path as the function of displacement to analyze feed rate, acceleration, jerk and contour error. Chen et al. [12] introduced contour error to feed rate scheduling and schedule feed rate in the contour error violate zone of tool path. A challenge for time-optimal approach is that it is difficult to solve optimal problems with many constrains, effectively and accurately since constraints are usually non-linear representation. To reduce the computational difficulty, linearization method was used in time-optimal method [13,14]. In other articles [15,16,17], linear programming was applied to discrete the whole movement process. Though these methods can improve computational efficiency, the computational burden is still large. After solving time-optimal problem, the feed rate profile is a set of scatters which cannot be applied to interpolation conveniently. Therefore, fitting feed rate with spline is necessary to obtain continue and smooth feed rate profile [8,13,14].

It is obviously that time-optimal approach can generate minimum time or near minimum time feed rate profile. However, some shortcomings still exist. One is computational complexity when solving the optimal problems and fitting feed rate profile. The other is that some problems about machining quality are difficult to be solved i. e. round off error [18]. Therefore, a number of feed rate scheduling methods have been developed using a variety of acceleration/deceleration (ACC/DEC) approach.

In this approach, the acceleration and deceleration process are designed in advance with the limitation of geometric and dynamic error. Usually, tool motion is a complex motion which contains multiple acceleration and deceleration processes. Thus, the first step is splitting integral machining process into several phases. Feed rate value will change drastically at the interval where curvature is large. These intervals can be defined as sharp corners [19] or critical zones [20]. Also, high-curvature points and C0 continue points can be determined as break-points which are used to split curve [21]. Other way is determining feed rate sensitive regions according to shape of feed rate profile [22,23]. Then, acceleration and deceleration process can be designed using ACC/DEC model. The most widely-used profiles in ACC/DEC approach are polynomial and trigonometric profiles. Cao. and Chang [24] used trapezoidal-velocity profile to design the acceleration and deceleration motion in their look ahead smoothly controlling algorithm. Smoothness of feed rate profile has better performance in machining quality. Erkorkmaz, K. and Altintas, Y. [25] applied trapezoidal-acceleration profile to generate smooth trajectory. In trapezoidal-acceleration profile, the whole process is divided into 7 phases. Generally, jerk-continues polynomial feed profile include 15 phases [26]. Furthermore, Wei et al. proposed a jerk-smooth feed rate profile in which whole process is divided into 31 phases [27]. For polynomial feed rate method, it can obtain near time-optimal feed rate profile since there are several adjustable parameters.

However, polynomial method is relatively difficult to implement because of more unknown parameters. For trapezoidal-acceleration profile, it contains 3 independent parameters which should be determined under symmetrical conditions. However, since the actual situation is usually asymmetrical, there are 5 independent parameters needed to be determined, as demonstrated in Fig.1. Therefore, trigonometric profile was also used in feed rate scheduling [21]. Lee et, al. [21] proposed a novel feed rate scheduling method contain two steps: splitting curve by break-points and designing ACC/DEC phase by sine curve. Nevertheless, machining efficiency was relatively low. In other articles [19,27,28], some feed rate scheduling method have been proposed based on initial sine-curve method. Wang et, al. [27] added linear term after trigonometric term so that feed rate profile was jerk-continues. Liu et, al. [19] designed jerk-continues feed rate profile similar to jerk-continues polynomial profile, and just replaced linear function with cosine function and abandoned constant non-zero phase in jerk profile. Huang and Zhu [28] interpolated parametric tool path using the sine series representation of jerk profile. It can be proved that its machining efficiency is higher than trigonometric profile. In fact, the low efficiency of trigonometric profile is derived from the representation of trigonometric function. Because of the lack of adjustable parameters, the shape of trigonometric profile is fixed and cannot achieve better efficiency.

In summary, time-optimal approach can be more efficient than ACC/DEC approach but there are more computation burdens. Furthermore, ACC/DEC approach can handle some special problems. On the other hand, polynomial method is more efficient compared with trigonometric method. However, trigonometric method is a succinct method so that it is convenient to be applied.

In order to guarantee the simplicity of the method and improve efficiency, a novel feed rate planning method based on Sigmoid function is proposed in this article. The method firstly obtains feed rate data with chord error constraint in the form of scatters. A feed rate detection method with chord error limitation is proposed. In this method, feed rate will be adjusted step by step, and adjusting coefficient will recompute in each step to ensure the accuracy. By scanning these scatters with method of two steps screening, some special points are selected to split curve into some blocks. For each block, there is only one acceleration, deceleration or constant feed rate process which is convenient to design feed rate profile. Then, a smooth feed rate profile is designed, and a time-optimal method is proposed to reach the minimum machining time. Compared with polynomial method, proposed method has less unknown parameters, and it has more excellent performance in term of machining efficiency than trigonometric method.

The rest of this paper is organized as follows. In Section 2, properties about Sigmoid functions are explained. The feed rate determining method with the chord error constraint and implementation are present in Section 3. In Section 4, a feed rate scheduling method based Sigmoid functions is proposed. In Section 5, simulations with two free-form NURBS curves are conducted to verify the feasibility and applicability of the proposed method, and the comparison with sine method is also presented in this section. Our conclusions and future work are summarized in Section 6.

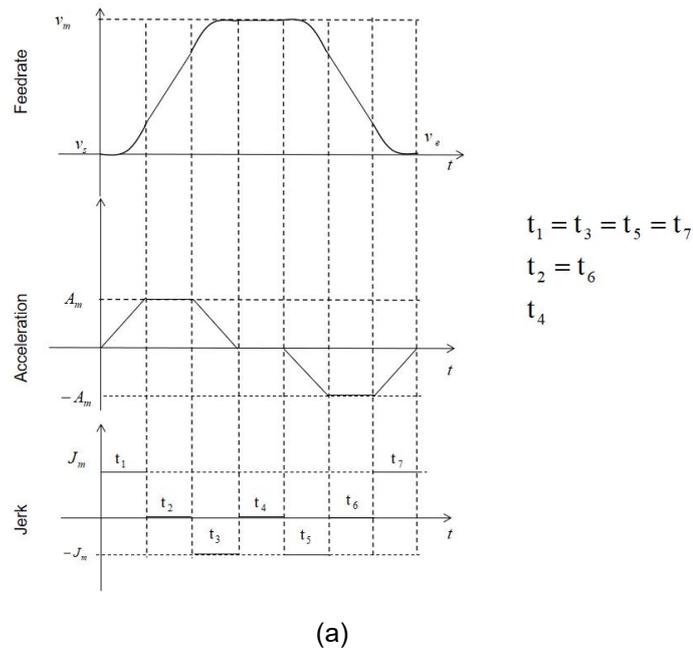

(a)

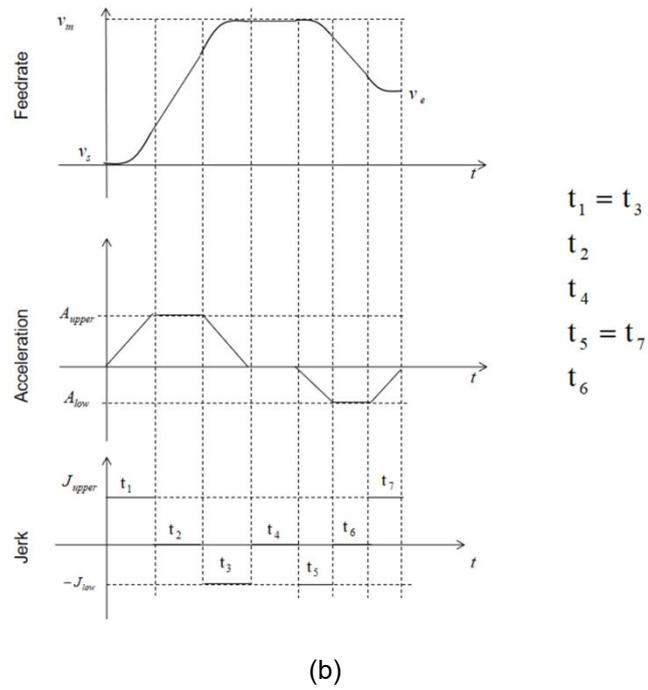

(b)

Fig. 1. (a) Trapezoidal-acceleration profile with symmetrical conditions, (b)Trapezoidal-acceleration profile with asymmetrical conditions

## 2. Sigmoid function and sine feed rate profile

2.1 Properties of Sigmoid function

Sigmoid function is widely used in machine learning as discrimination function to predict and classify data because it is monotone differentiable, arbitrary order differentiable and S-shape. Initially, unit-step function is charged with the task of classification and prediction in machine learning. However, discontinuous and non-differentiable characters of unit-step function are not convenient when some numerical optimization algorithms are used. Therefore, Sigmoid function instead of unit-step function is applied in the model of machine learning. The Sigmoid function is defined by the formulation (1)

$$f(x) = \frac{1}{1+e^{-x}} \qquad (1)$$

Domain of the function: $(-\infty, +\infty)$

Range of the function: $(0,1)$

And: $\lim_{x \to -\infty} f(x) = 0$  $\lim_{x \to +\infty} f(x) = 1$  Its derived function is the form of a product which include two function about $f(x)$. As the equation (2),

$$f'(x) = f(x)(1-f(x)) \qquad (2)$$

The graph of Sigmoid function is in S curve form which is shown in the Fig. 2(a). and Fig. 2(b) shows the shape of its derived function.

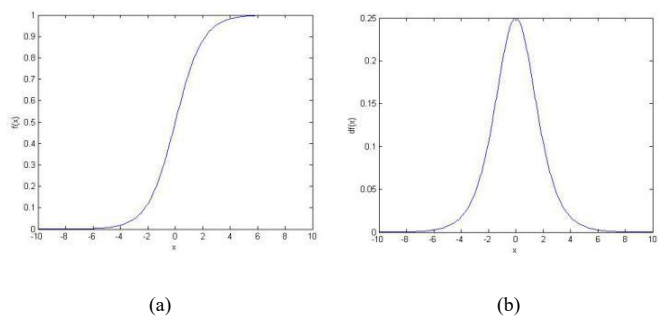

(a)        (b)

Fig.2. (a) Graph of Sigmoid function;(b) Graph of derived function

Define two functions about Sigmoid function defined as equations (3) and (4):

$$f_1(x) = \frac{x}{2f(x)-1} = \frac{x(1+e^{-x})}{1-e^{-x}} \tag{3}$$

$$f_2(x) = \frac{x^2}{2f(x)-1} = \frac{x^2(1+e^{-x})}{1-e^{-x}} \tag{4}$$

Function (1) is an increasing function. For deceleration, a function $p(x)$ that is similar to $f(x)$ is adopted. It can be defined as equation (5):

$$p(x) = f(-x) = \frac{1}{1+e^x} \tag{5}$$

$p(x)$ in negative infinity is 1. $p(x)$ in positive infinity is 0.

$$\lim_{x \to +\infty} p(x) = 0 \quad \text{and} \quad \lim_{x \to -\infty} p(x) = 1$$

Similarly, its derived function can be determined, as equation (6):

$$p'(x) = p(x)(p(x)-1) \tag{6}$$

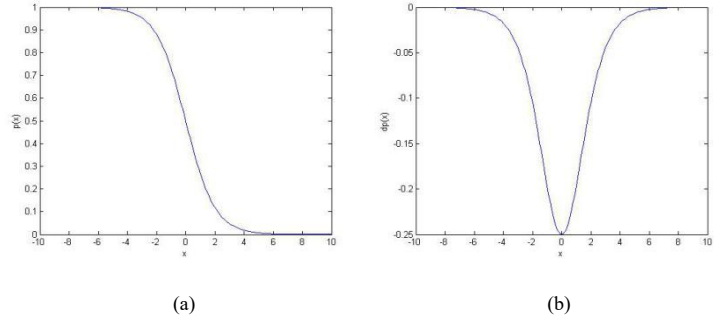

(a)          (b)

Fig. 3. (a) Graph of function p(x); (b) Graph of the derivation function of p(x)

Two functions $p_1(x), p_2(x)$ can be defined as the same approach as the $f_1(x), f_2(x)$.

$$p_1(x) = \frac{x}{2p(x)-1} = \frac{x(1+e^x)}{1-e^x} \tag{7}$$

$$p_2(x) = \frac{x^2}{2p(x)-1} = \frac{x^2(1+e^x)}{1-e^x} \tag{8}$$

There are some properties which can be used in following sections:

1, When x is greater than zero, $f_1(x), f_2(x), p_1(x), p_2(x)$ are monotonic functions. $f_1(x), f_2(x)$ are monotonic increasing functions. And, $p_1(x), p_2(x)$ are monotonic decreasing functions.

2, When x is in the interval $[0,+\infty)$, the ranges of $f_1(x), f_2(x)$, $p_1(x)$ and $p_2(x)$ are $(2,+\infty), (0,+\infty), (-\infty,-2)$ and $(-\infty,0)$, respectively.

3, The maximum value of $f'(x)$ is 1/4 and the minimum value of $p'(x)$ is -1/4.

In kinematic knowledge, acceleration is the derivative of velocity and jerk can be obtained by taking derivative of acceleration. So, second derivatives of $f(x)$ or $p(x)$ are necessary which are shown in equations (9) and (10).

$$f''(x) = 2f^3(x) - 3f^2(x) + f(x) \qquad (9)$$

$$p''(x) = 2p^3(x) - 3p^2(x) + p(x) \qquad (10)$$

4, The value of $f''(x)$ is maximum when $f(x) = 0.5 - \sqrt{3}/6$ and the value of $p''(u)$ is minimum when $p(x) = 0.5 + \sqrt{3}/6$. The graphs of function (3), (4), (7) and (8) are shown in Fig. 4. (a), (b), (c) and (d), respectively.

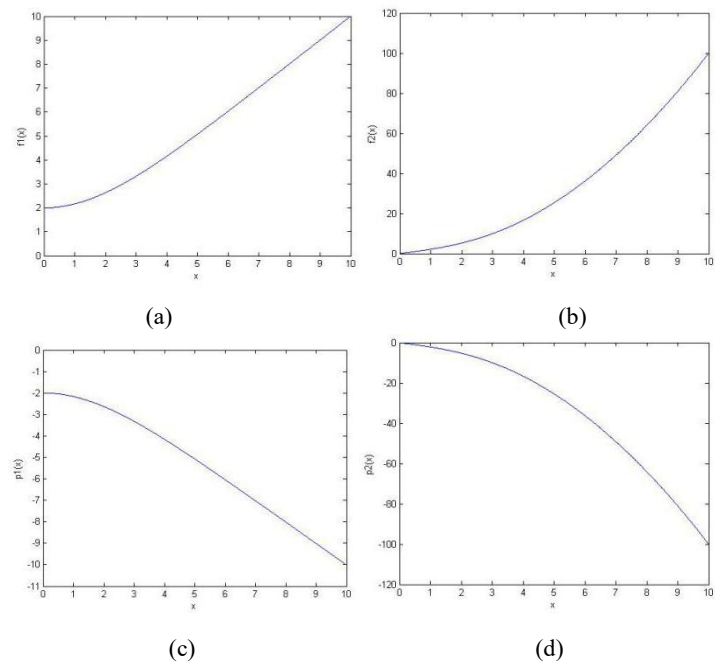

(a)     (b)

(c)     (d)

Fig. 4. (a) Graph of function (3); (b) Graph of function (4); (c) Graph of function (7); (d) Graph of function (8)

2.2 Sine feed rate profile

Since sine-curve velocity profile is smooth curve and more concise than polynomial velocity profile, it can be used to generate the feed rate profile and its velocity

equation is given as equation (11) in [21]:

$$v(t) = \frac{v_e - v_s}{2}\left[\sin \pi\left(\frac{t}{T} - \frac{1}{2}\right) + 1\right] + v_s, \quad 0 \leq t \leq T \tag{11}$$

Where $v_s$ and $v_e$ denote the start and end feed rate. $T$ denote the time from start to end. Differentiating equation (11) yields the acceleration equation as (12),

$$A(t) = \frac{v_e - v_s}{2}\frac{\pi}{T}\cos \pi\left(\frac{t}{T} - \frac{1}{2}\right), \quad 0 \leq t \leq T \tag{12}$$

Differentiating equation (12), we can obtain the jerk equation,

$$J(t) = -\frac{v_e - v_s}{2}\left(\frac{\pi}{T}\right)^2 \sin \pi\left(\frac{t}{T} - \frac{1}{2}\right) \tag{13}$$

The limits of tangent acceleration and jerk can be determined as equations (13) and (14):

$$|A(t)| = \left|\frac{v_e - v_s}{2}\frac{\pi}{T}\cos \pi\left(\frac{t}{T} - \frac{1}{2}\right)\right| \leq \left|\frac{v_e - v_s}{2}\right|\frac{\pi}{T} \leq A_m \tag{14}$$

$$|J(t)| = \left|\frac{v_e - v_s}{2}\left(\frac{\pi}{T}\right)^2 \sin \pi\left(\frac{t}{T} - \frac{1}{2}\right)\right| \leq \left|\frac{v_e - v_s}{2}\right|\left(\frac{\pi}{T}\right)^2 \leq J_m \tag{15}$$

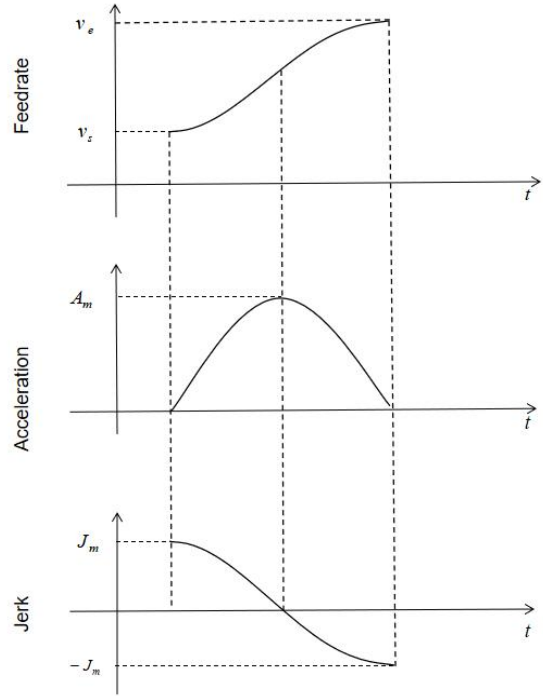

Fig. 5. Sine feed rate profile

The feed rate, acceleration and jerk profile are shown in Fig. 5. According to equations (11), (12) and (13), the feed rate curve and acceleration curve can be guaranteed to be continuous but jerk curve is discontinuous because the value of jerk is not zero at start and end point.

## 3. Feed rate generation under the chord error constraint

To generate feed rate profile of given parametric curve, an off-line process for pre-interpolation and feed rate data scanning, which is aimed to determine the feed rate data of curve, is developed. In the stage of pre-interpolation, the feed rate data of the whole curve is determined in the form of scatters. By using an approximate ratio relationship, feed rate data which satisfies the chord error constraint is obtained to prepare for the following stage. In the next stage, the breaking points are determined by two steps. Above all, screening factor is computed at every feed rate point firstly. According to the screening factor and given standard value, the candidates of breaking point can be selected when the screening is greater than the standard value. In candidate points, there are some noise points which are not compatible with the definition of breaking point. Then, a symbolic function is defined to distinguish the noise points from the true breaking points. The feed rate curve can be split into some sub-curves via these points. In order to record these interval segments, a structure called block is defined containing the start position parameter $u_s$, end position parameter $u_e$, start feed rate $v_s$, end feed rate $v_e$, time $T$, shape parameter $s$ and displacement $L$. The displacement is computed by using numerical integration method, and the shape parameter and time will be computed in the section 4.

3.1 Chord error constraint

In parametric interpolation, the desired trajectory can be described as the kinds of parametric curves, such as Bezier curve, B-spline curve, Hermite curve, and NURBS curve. All of these can be present as $C(u)$. In parametric interpolation, data sampling method is a common method. With this method, the main stage is recurrently computing the next sampling point based on current point using the fixed sampling time $T_s$. After computing all sampling points, the real tool path is composed of the links between the adjacent sampling points within the fixed $T_s$. Suppose the relationship between parameter $u$ and time $t$ can be expressed as $u(t)$. During computing sampling points, the parameters of adjacent points satisfy $u_{i+1} = u_i + \Delta(u_i)$, where $u_i$ is the parameter value of the current sampling point, and $u_{i+1}$ is the parameter value of the next sampling point, while $\Delta(u_i)$ is the incremental value within a sampling time. Then, a two order Taylor expansion [29] can be used to describe the relationship between parameter u and sampling time $T_s$. As the equation (16):

$$u_{i+1} = u_i + \frac{v_i T_s}{\left\| \frac{dC(u)}{du} \right\|_{u=u_i}} - \frac{(\frac{dC(u)}{du} \frac{d^2C(u)}{du^2})}{2 \bullet \left\| \frac{dC(u)}{du} \right\|^4_{u=u_i}} v_i^2 T_s^2 \tag{16}$$

where $v_i$ is the feed rate value within a sampling period.

Two adjacent sampling points are connected by straight line. The parametric curve is approximated by these short straight lines, so the chord error is inevitable if the curvature of the curve does not equal with 0, as shown in Fig. 6(a).

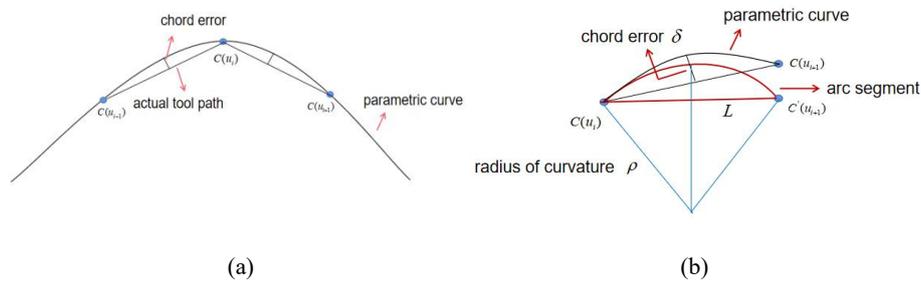

Fig. 6. (a) Parametric curve approximated by two short straight lines, (b) Schematic diagram of chord error.

The chord error is defined as the maximum distance between the parametric curve and actual tool path. Because the distance between two adjacent sampling points is very close, the radius of curvature at two points can be considered equal. So that an arc segment is used to approximate this curve between the two adjacent points [30], as shown in Fig. 6(b).

From the approximation model, the equation about chord error and chord length can be obtained [30] as equation (17):

$$L = 2\sqrt{\delta(2\rho - \delta)} \tag{17}$$

Where L is chord length, δ is chord error and ρ is radius of curvature. Because the chord length is the distance of actual tool path, an equal about feed rate and chord length is obtained as equation (18):

$$L = vT_s \tag{18}$$

Substitute equation (18) into equation (17), and we can get equation about feed rate and chord error (19):

$$v = \frac{2}{T_s}\sqrt{\delta(2\rho - \delta)} \tag{19}$$

Since the value of the chord error is very small, the quadratic term of chord error can be ignored. Then:

$$v \sim \sqrt{\delta}$$

If a chord error tolerance $\delta_m$ is given, the feed rate under the chord error constraint $v_m$ is obtained by equation (20):

$$v_m = \sqrt{\delta_m/\delta}\, v \tag{20}$$

Denote the coefficient of v as τ called feed rate adjusting coefficient. However, the arc model is an approximation model and the quadratic term of chord error in equation (19) is ignored. Consequently, recurrent adjustment for feed rate is necessary to satisfy chord error. There is a strategy to adjust feed rate recurrent.

Step 1, according to the given parameter of current point $u_i$, computing the parameter of next sampling $u'_{i+1}$;

Step 2, computing the chord error $\delta$ between the two points $C(u_i)$ and

$C(u'_{i+1})$, and comparing with the chord error tolerance $\delta_m$. If $\delta > \delta_m$, then go to step 3. Else $\delta \leq \delta_m$, go to step 4;

Step 3, adjusting the feed rate just as the equation (20). Then go to the Step 1;

Step 4, let $u_{i+1} = u'_{i+1}$.

To determine the feed rate profile about whole parametric curve, a pre-interpolation process is proposed. In this process, two order Taylor expansion is used to compute the parameter u. Then the above strategy is applied to compute the feed rate at the current point. When the parameter is equal to 1, the process ends. The whole process is shown in the follow flow chart Fig. 7.

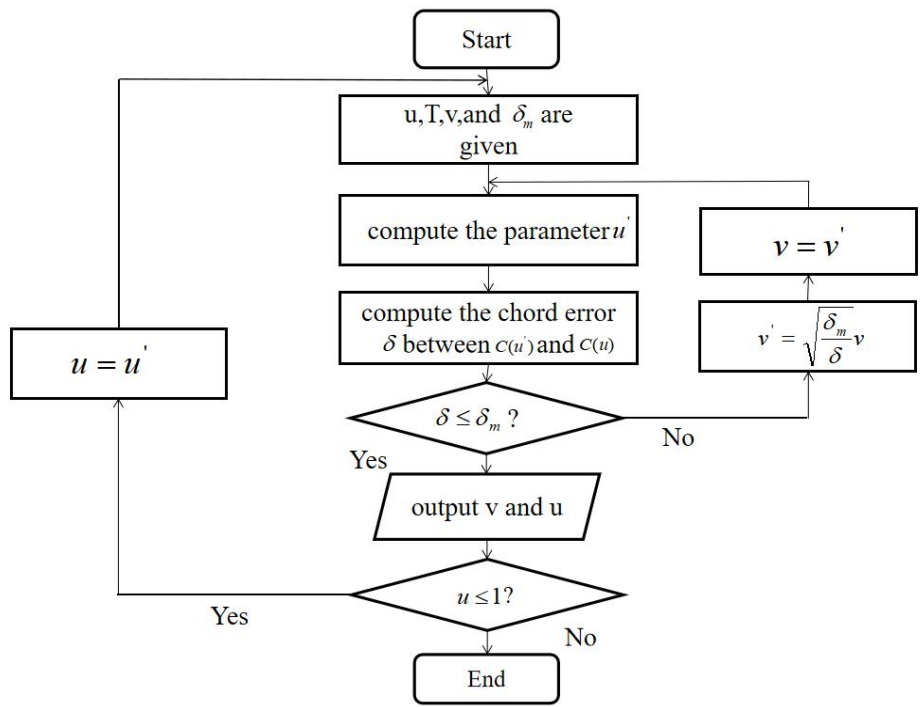

Fig. 7: Flow chart of pre-interpolation process

3.2 Feed rate profile splitting

In order to simplify the process of feed rate scheduling, feed rate profile, which is determined by the pre-interpolation process mentioned above, should be split into several segments. There are three types of segments: acceleration segment, deceleration segment and constant feed rate segment after splitting curve.

These segments are determined by breaking points which are the points where the trend of feed rate changes is different on both sides on the curve. An example of breaking point is shown in the Fig. 8. (a). On the left of the point A, it is acceleration segment. On the right of the point A, it is deceleration segment. Therefore, point A is the breaking point.

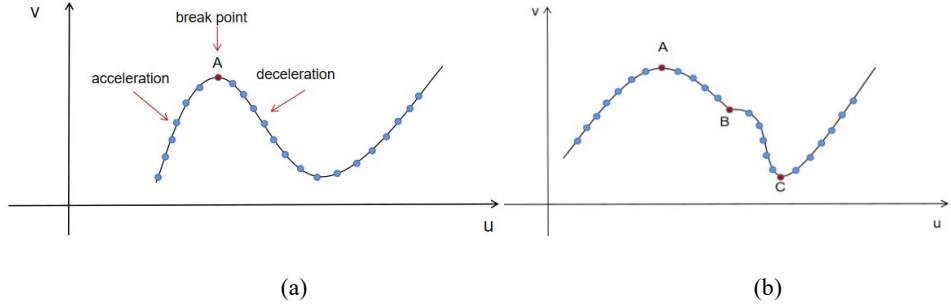

(a) (b)

Fig. 8:(a) Breaking point in the curve; (b) Noise point and breaking points in the curve

Therefore, if the breaking points are determined, the feed rate profile can be divided into several segments mentioned above. These points can be determined through two steps. The first step is computing the screening factor. Assume that there are three points of feed rate profile, $(u_{i-1}, v_{i-1})$, $(u_i, v_i)$ and $(u_{i+1}, v_{i+1})$.

Denote screening factor at $u_i$ as $\mu_i$, and it can be computed by the following equation (21):

$$\mu_i = \left| \frac{v_{i+1} - v_i}{u_{i+1} - u_i} - \frac{v_i - v_{i-1}}{u_i - u_{i-1}} \right| \qquad (21)$$

When the screening factor is greater than a given standard value $\mu_s$, the point is a candidate of breaking point. However, there are some points which are not breaking points but their screening factors are also greater than $\mu_s$, as shown in Fig.8(b). The point A and C are breaking point, but the screening factor at point B can also be greater than $\mu_s$. Obviously, point B is not breaking point. So, the second step is necessary to determine the true breaking points.

A series of points $(u_i, v_i), 0 \leq i \leq n$ which are the candidates of breaking points are obtained after first step. In the second step, the points, which constitute monotonous sequence of points with two adjacent points, are deleted. In order to separate the false inflection point from the true inflection point, a symbolic function can be calculated as the equation (22):

$$w(i) = \begin{cases} 1, (v_{i+1}-v_i)(v_i-v_{i-1}) > 0 \\ -1, (v_{i+1}-v_i)(v_i-v_{i-1}) < 0 \end{cases}, 0 \le i \le n \tag{22}$$

Where $v_{i+1}$, $v_i$ and $v_{i-1}$ are the feed rate of points $(u_{i+1}, v_{i+1})$, $(u_i, v_i)$ and $(u_{i-1}, v_{i-1})$ respectively. When $w(i)$ is equal to -1, the point $(u_i, v_i)$ is the breaking point which satisfy the requirements.

Through the two steps, the breaking points are determined. Then, the feed rate profile can be divided into some segments. Feed rate between the two adjacent breaking points may be increased, decreased or unchanged. In the first step, it is noticed that the standard value is small enough to ensure that the real breaking points can be selected.

Then, the feed rate curve can be divided into several sub-curve. Each sub-curve contains the process of feed rate change, and each sub-curve needs to be recorded. Hence, a structure named as block, which consists of start position parameter, end position parameter, start feed rate, end feed rate, time, displacement and shape parameter, is defined.

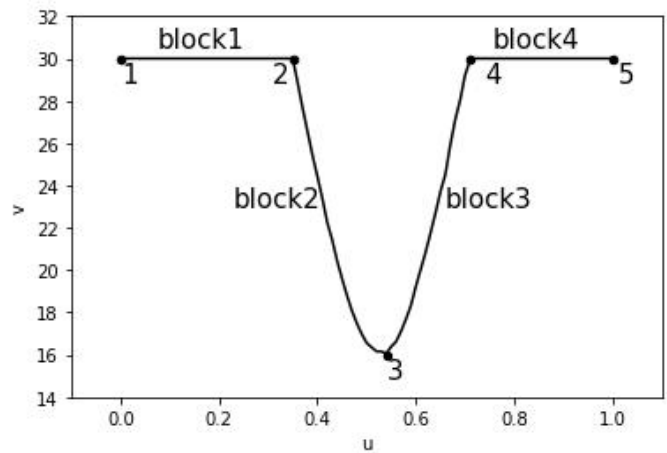

Fig. 9. Feed rate curve split into 4 blocks.

As shown in the Fig. 9, point 2, point 3 and point 4 are breaking points. Point 1 is the start point which the parameter of is equal to 0. Similarly, point 5 is the end point which the parameter of is equal to 1. The curve is the feed rate curve of a spatial cubic curve obtained by the strategy mentioned above. These points split the feed rate curve into 4 blocks. The displacement in the blocks is acquired by computing integral.

$$L = \int_{u_s}^{u_e} ds = \int_{u_s}^{u_e} \sqrt{x'^2(u) + y'^2(u) + z'^2(u)} \, du \tag{23}$$

Where $x'(u)$, $y'(u)$ and $z'(u)$ are the first-order derivative of parameter curve coordinate components $x(u)$, $y(u)$ and $z(u)$ respectively.

The other elements, namely shape parameter and time will be computed in the next section.

## 4. Feed rate scheduling based on Sigmoid function

When the process of feed rate data scanning is finished, some elements of block are obtained and stored. In the stage of feed rate scheduling, the aim is to construct a jerk-limited feed rate profile for each NURBS block through the scanning data. The next important stage is to construct a jerk-limited feed rate profile for each block. For one block, the velocity of start and end points are known but the velocity of other points between the start and end points are unknown. For the block where the $v_s$ is equal to $v_e$, when the time $t$ is $t_0$, the velocity can be denoted as $v(t_0) = v_s$ or $v(t_0) = v_e$. Then, for other block, the velocity of point where the time is $t_0$, can be computed by feed rate scheduling method and denoted as $v(t_0)$.

4.1 velocity profile based on Sigmoid functions

Because of the shape like the letter S, the Sigmoid function can be used to design the velocity profile. For a block which represents a accelerated process, the function (1) mentioned in the Section 2 can be used.

Suppose that a section symmetrical about the origin is intercepted from the real number axis, which is recorded as $[-s, s]$. Then, the function range is $[f(-s), f(s)]$. Where the parameter s is the shape parameter.

There are two maps that should be constructed.

$$g_1 : [0, T] \mapsto [-s, s]$$

$$g_2 : [f(-s), f(s)] \mapsto [v_s, v_e]$$

In fact, both of maps are linear functions. Just like equations (25) and (26):

$$g_1 = \frac{2s}{T} t - s \tag{25}$$

$$g_2 = \frac{v_e - v_s}{f(s) - f(-s)} (f(g_1) - f(-s)) + v_s \tag{26}$$

The velocity equation is given as following equation (27):

$$v(t) = \frac{v_e - v_s}{f(s) - f(-s)} (f(\frac{2s}{T} t - s) - f(-s)) + v_s \tag{27}$$

Where T is the time in block. Time T and displacement L satisfy the integral equation(28):

$$L = \int_0^t v(\tau) d\tau \tag{28}$$

However, due to the symmetry of the velocity function, the equation (28) can be simplified to equation (29):

$$L = \frac{(v_s + v_e)}{2} T \tag{29}$$

Then, differentiating equation (27) yields the acceleration equation (30),

$$A(t) = \frac{2s}{T} \frac{v_e - v_s}{f(s) - f(-s)} f(\frac{2s}{T} t - s)(1 - f(\frac{2s}{T} t - s)) \tag{30}$$

Differentiating equation (30), one obtains the jerk equation (31):

$$J(t) = (\frac{2s}{T})^2 \frac{v_e - v_s}{f(s) - f(-s)} (f(\frac{2s}{T} t - s)(1 - f(\frac{2s}{T} t - s)) - 2f^2(\frac{2s}{T} t - s)(1 - f(\frac{2s}{T} t - s))) \tag{31}$$

The premise of these equations is that the process is accelerated. For the deceleration process, function p(x) is applied. The velocity equation, acceleration

equation and jerk equation are given as (32), (33) and (34):

$$v(t) = \frac{v_e - v_s}{p(s) - p(-s)}(p(\frac{2s}{T}t - s) - p(-s)) + v_s \tag{32}$$

$$A(t) = \frac{2s}{T}\frac{v_e - v_s}{p(s) - p(-s)} p(\frac{2s}{T}t - s)(p(\frac{2s}{T}t - s) - 1) \tag{33}$$

$$J(t) = (\frac{2s}{T})^2 \frac{v_e - v_s}{p(s) - p(-s)}(p(\frac{2s}{T}t - s)(p(\frac{2s}{T}t - s) - 1) - 2p^2(\frac{2s}{T}t - s)(p(\frac{2s}{T}t - s) - 1)) \tag{34}$$

There are limits of tangent acceleration and jerk:

$$\begin{cases} |A(t)| \leq A_m \\ |J(t)| \leq J_m \end{cases} \tag{35}$$

Using the properties 3 and 4 in Section2, the inequality can be further reduced to the following inequality (36):

$$\begin{cases} \dfrac{s}{2f(s)-1}\dfrac{v_e - v_s}{T}\dfrac{1}{2} \leq A_m \\ \dfrac{4s^2}{2f(s)-1}\dfrac{v_e - v_s}{T^2}\left|2k^3 - 3k^2 + k\right| < J_m, k = 0.5 - \sqrt{3}/6 \end{cases} \tag{36}$$

Where the $A_m$ and $J_m$ are the maximum acceleration and jerk which the machine tool can provide, respectively. For the deceleration process, the function $f(s)$ is replaced with $p(x)$.

4.2 Feed rate smoothing strategy

Although the Sigmoid function is smooth and higher order differentiable, the derivative value at s or -s is not zero. Hence, the feed rate profile is only $c^0$ continuous. To obtain continuous acceleration profile, a feed rate smoothing strategy is applied.

For one block, the process which block present is divided into 3 parts, $[0, T/3]$, $[T/3, 2T/3]$, $[2T/3, T]$. In the parts $[0, T/3]$ and $[2T/3, T]$, the feed

rate curve is replaced by a cubic curve as equation (37):

$$\bar{v}(t) = \begin{cases} a_1 t^3 + a_2 t^2 + a_3 t + a_4 & t \in [0, T/3] \\ b_1(T-t)^3 + b_2(T-t)^2 + b_3(T-t) + b_4 & t \in [2T/3, T] \end{cases} \quad (37)$$

Then, the acceleration and jerk profile can be obtained as the equations (38) and (39):

$$\bar{A}(t) = \begin{cases} 3a_1 t^2 + 2a_2 t + a_3 & t \in [0, T/3] \\ -3b_1(T-t)^2 - 2b_2(T-t) - b_3 & t \in [2T/3, T] \end{cases} \quad (38)$$

$$\bar{J}(t) = \begin{cases} 6a_1 t + 2a_2 & t \in [0, T/3] \\ 6b_1(T-t) + 2b_2 & t \in [2T/3, T] \end{cases} \quad (39)$$

In order to obtain the acceleration continue curve, there are some conditions that need to meet equations (40):

$$\bar{v}(0) = v_s \quad \bar{v}(T/3) = v(T/3) \quad \bar{v}(2T/3) = v(2T/3) \quad \bar{v}(T) = v_e \quad (40)$$

$$\bar{A}(0) = 0 \quad \bar{A}(T/3) = A(T/3) \quad \bar{A}(2T/3) = A(2T/3) \quad \bar{A}(T) = 0$$

Because of the kinematic constraints, the inequalities (35) are necessary. Through the given conditions, the parameters in the cubic function can be solved as equations (41).

$$\begin{aligned} a_4 &= v_s \\ b_4 &= v_e \\ a_3 &= b_3 = 0 \\ a_2 &= -b_2 = \frac{27(v(T/3) - v_s)}{T^2} - \frac{3A(T/3)}{T} \\ a_1 &= -b_1 = \frac{9A(T/3)}{T^2} - \frac{54(v(T/3) - v_s)}{T^3} \end{aligned} \quad (41)$$

Then, an acceleration continuous feed rate profile is built, as shown in the Fig. 10. For one block, the whole process is split into three parts. In the first and third parts, cubic function is used to build the feed rate profile. In the second, the feed rate profile is based on the Sigmoid function that has undergone two compound changes.

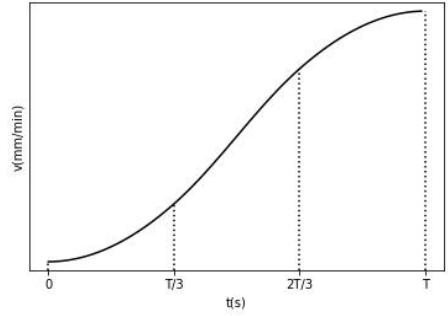

Fig. 10. Feed rate profile consist of 3 section

A fact is that when the shape parameters $s$ take appropriate value, this method is more efficient than sine-curve feed rate scheduling method.

4.3, Time-optimal adjustment of feed rate

From the Section 3, the feed rate data is divided into some blocks by breaking points. Although the feed rate meets the geometric error constraint, not all processes can meet the kinematic constraint. For some blocks which have short displacement and large difference between start and end feed rate, their acceleration and jerk values may exceed the limits of acceleration $A_m$ and jerk $J_m$. Hence, the adjustment method about these blocks is important to ensure that acceleration and jerk can satisfy the kinematic limits. There are four situations need to be considered as shown in Fig.11. The first is the process whose feed rate is increased firstly and then constant, in the Fig.11(a). The second situation is just the opposite of first situation as shown in the Fig.11(b). The next is the process whose feed rate is increased firstly and then decreasing, in the Fig.11(c). The final one is the process which contains the constant feed rate phase, in the Fig.11(d).

For the first two situations, displacements of constant feed rate blocks can be decreased. When displacements of constant feed rate blocks become 0, feed rate in constant blocks need to adjusting, just as shown in Fig.11(a) and (b).

For the third situation, when the shape parameter $s$ is fixed, the only parameters of block can be modified is the feed rate at the start or end point. In the Fig. 11(c), from $u_{i-1}$ to $u_i$, the process is the $block_{i-1}$. And, from $u_i$ to $u_{i+1}$, the process is the $block_i$. The feed rate $v_1$, $v_2$ and $v_3$ are the feed rate at $u_{i-1}$,

$u_i$ and $u_{i+1}$ respectively. It is obviously that reducing the feed rate $v_2$ is an effective method. The feed rate $v_2'$ is the feed rate at $u_i$ which is adjusted.

Machining efficiency is a significant factor that is necessary to considered. The whole time can be expressed as equation (42).

$$T = T_1 + T_2 = \frac{2L_1}{v_1 + v_2} + \frac{2L_2}{v_2 + v_3} \tag{42}$$

Where $L_1$ and $L_2$ are the displacements of $block_{i-1}$ and $block_i$ respectively. Then, it becomes an optimal problem with the constraints (43).

$$\min \quad \frac{L_1}{v_1 + v_2} + \frac{L_2}{v_2 + v_3} \tag{43}$$

$$\begin{cases} |A_i(t)| \leq A_m, i=1,2 \\ |J_i(t)| \leq J_m, i=1,2 \\ |\overline{A}_i(t)| \leq A_m, i=1,2 \\ |\overline{J}_i(t)| \leq J_m, i=1,2 \end{cases}$$

Substitute equations (36) and (41) into (43):

$$\frac{v_2 - v_1}{f(s) - f(-s)} \frac{s(v_2 + v_1)}{L_1} \frac{1}{4} \leq A_m$$

$$\frac{v_3 - v_2}{p(s) - p(-s)} \frac{s(v_3 + v_2)}{L_2} \frac{1}{4} \leq A_m$$

$$\frac{v_2 - v_1}{f(s) - f(-s)} \frac{s^2(v_2 + v_1)^2}{L_1^2} \lambda_1 \leq J_m \tag{44}$$

$$\frac{v_3 - v_2}{p(s) - p(-s)} \frac{s^2(v_2 + v_3)^2}{L_2^2} \lambda_2 \leq J_m$$

$$\left| \frac{a_{i,2}^2}{3a_{i,1}} \right| \leq A_m \quad i = 1,2$$

$$\max(|2a_{i,2}|, |6a_{i,1} + 2a_{i,2}|) \leq J_m \quad i = 1,2$$

Where the parameter $\lambda_1$ is the value of equation (9) in Section 2 when the k is

$0.5 - \sqrt{3}/6$ in the $block_{i-1}$, and the parameters $a_{1,1}$ and $a_{1,2}$ are the coefficients of cubic term and quadratic term of first cubic function respectively in the $block_{i-1}$. Similarly, the parameter $\lambda_2$ is the value of equation (10) in Section 2 when the k is $0.5 + \sqrt{3}/6$. The parameters $a_{2,1}$ and $a_{2,2}$ are the coefficients of cubic term and quadratic term of first cubic function respectively in the $block_i$.

For this optimal problem, the constraints are polynomials about $v_2$. It is obviously that the value of $v_2$ should be as large as possible. Then, for each inequality, when the equal sign is established, the value of the $v_2$ reaches the maximum value. The minimum value of these maximum values is the solution of the optimal problem.

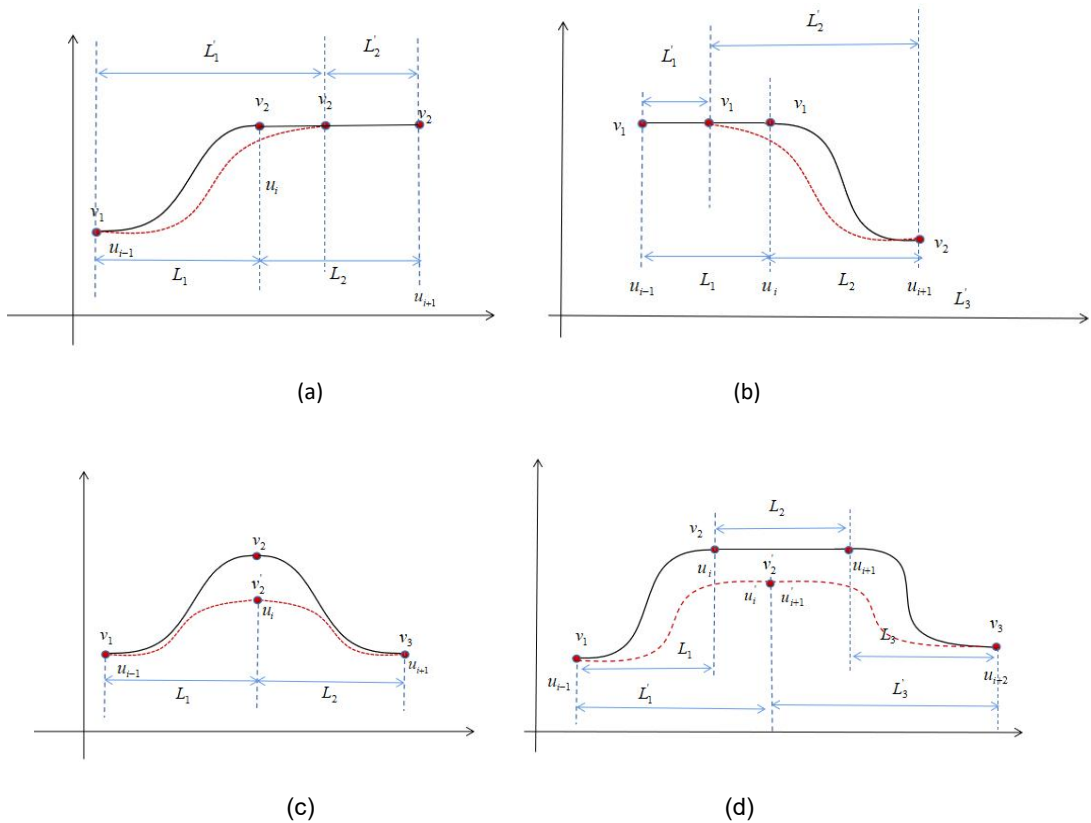

Fig. 11. (a) Feed rate adjusting for two blocks with acceleration and constant feed rate phase, (b) Feed rate adjusting for two blocks with constant feed rate and deceleration phase, (c) Feed rate adjusting for two blocks with no constant feed rate phase, (d) Feed rate adjusting for three blocks.

For the last situation, as shown in the Fig. 11(d), there are three process. From

$u_{i-1}$ to $u_i$, the $block_{i-1}$ is accelerated and the displacement is $L_1$. From $u_i$ to $u_{i-1}$, it is a constant feed rate phase and its displacement can be denoted as $L_2$. Similarly, from $u_{i+1}$ to $u_{i+2}$, the $block_{i+2}$ is decelerating whose displacement is remarked as $L_3$. To ensure the demand of chord error and reduce machining time as much as possible, $L_1$, $L_3$ and the feed rate value of constant feed rate phase are need to be adjusted.

In the last situation, the machining time can be expressed as equation (45):

$$T = T_1 + T_2 + T_3 = \frac{2L_1}{v_1 + v_2} + \frac{L_2}{v_2} + \frac{2L_3}{v_2 + v_3} \tag{45}$$

Where $v_1$ and $v_3$ are the feed rate at $u_{i-1}$ and $u_{i+2}$ respectively. Then, an optimal problem also can be obtained.

$$\text{Min} \quad \frac{L_1}{v_1 + v_2} + \frac{L_2}{v_2} + \frac{L_3}{v_2 + v_3} \tag{46}$$

$$\begin{cases} |A_i(t)| \leq A_m, i = 1,3 \\ |J_i(t)| \leq J_m, i = 1,3 \\ |\overline{A}_i(t)| \leq A_m, i = 1,3 \\ |\overline{J}_i(t)| \leq J_m, i = 1,3 \\ L_1 + L_2 + L_3 = L(\text{constant}) \end{cases}$$

Substitute equations (36) and (41) into (46):

$$\frac{v_2 - v_1}{f(s) - f(-s)} \frac{s(v_2 + v_1)}{L_1} \frac{1}{4} \leq A_m$$

$$\frac{v_3 - v_2}{p(s) - p(-s)} \frac{s(v_3 + v_2)}{L_3} \frac{1}{4} \leq A_m$$

$$\frac{v_2 - v_1}{f(s) - f(-s)} \frac{s^2(v_2 + v_1)^2}{L_1^2} \lambda_1 \leq J_m$$

$$\frac{v_3 - v_2}{p(s) - p(-s)} \frac{s^2(v_2 + v_3)^2}{L_3^2} \lambda_2 \leq J_m \tag{47}$$

$$\left|\frac{a_{i,2}^2}{3a_{i,1}}\right| \leq A_m \quad i = 1,3$$

$$\max(|2a_{i,2}|, |6a_{i,1} + 2a_{i,2}|) \leq J_m \quad i = 1,3$$

$$L_1 + L_2 + L_3 = L = \int_{u_{i-1}}^{u_{i+2}} \sqrt{x'^2 + y'^2 + z'^2} \, du$$

For the optimal problem, if the feed rate $v_2$ is fixed, the longer the phase with constant feed rate, the shorter the time. For the constraints of acceleration, these inequalities can be expressed in the same form.

$$(v_2 - v_1)(v_2 + v_1)\mu_1 - A_m L_1 \leq 0 \tag{48}$$

$$(v_2 - v_3)(v_3 + v_2)\mu_1 - A_m L_3 \leq 0 \tag{49}$$

$$(v_2 - v_1)(v_2 + v_1)\mu_2 - A_m L_1 \leq 0 \tag{50}$$

$$(v_2 - v_3)(v_2 + v_3)\mu_2 - A_m L_3 \leq 0 \tag{51}$$

$$\mu_1 = \frac{1}{4} \frac{s}{f(s) - f(-s)} \tag{52}$$

$$\mu_2 = \frac{1}{2(f(s) - f(-s))} \left| \frac{81q^2 + 4s^2 p^2 - 36spq}{6sp - 54q} \right| \tag{53}$$

$$q = f(-\frac{s}{3}) - f(-s) \quad p = f(-\frac{s}{3})(1 - f(-\frac{s}{3}))$$

Parameter $\mu_1$ and $\mu_2$ reflect the strictness of the constraints. For example, if $\mu_1 > \mu_2$, the constraints (48) and (49) are stricter than constraints (50) and (51). In other words, if constraints (48) and (49) have been met, then conditions (50) and (51) must also be met.

In the same way, the constraints about jerk can be expressed as the follow form:

$$(v_2 - v_1)(v_2 + v_1)^2 \mu_3 - L_1^2 J_m \leq 0 \tag{54}$$

$$(v_2 - v_3)(v_2 + v_3)^2 \mu_3 - L_3^2 J_m \leq 0 \tag{55}$$

$$(v_2 - v_1)(v_2 + v_1)^2 \mu_4 - L_1^2 J_m \leq 0 \tag{56}$$

$$(v_2 - v_3)(v_2 + v_3)^2 \mu_4 - L_3^2 J_m \leq 0 \tag{57}$$

$$(v_2 - v_1)(v_2 + v_1)^2 \mu_5 - L_1^2 J_m \leq 0 \tag{58}$$

$$(v_2 - v_3)(v_2 + v_3)^2 \mu_5 - L_3^2 J_m \leq 0 \tag{59}$$

$$\mu_3 = \frac{s^2}{4(f(s) - f(-s))} \lambda_1 \tag{60}$$

$$\mu_4 = \frac{1}{4(f(s) - f(-s))} |54q - 12sp| \tag{61}$$

$$\mu_5 = \frac{1}{4(f(s) - f(-s))} |24sp - 54q| \tag{62}$$

As the same way, constraints about jerk which is strictest will be selected from the constraints (54), (55), (56), (57), (58) and (59) in the $block_{i-1}$. Therefore, the original optimization problem will be simplified to the form of (63):

$$\text{Min} \quad \frac{L_1}{v_1 + v_2} + \frac{L_2}{v_2} + \frac{L_3}{v_2 + v_3} \tag{63}$$

$$(v_2 - v_1)(v_2 + v_1)^2 \mu_m - L_1^2 J_m \leq 0$$

$$(v_2 - v_3)(v_2 + v_3)^2 \mu_m - L_3^2 J_m \leq 0$$

$$(v_2 - v_1)(v_2 + v_1) \mu_n - L_1 A_m \leq 0$$

$$(v_2 - v_3)(v_2 + v_3) \mu_n - L_3 A_m \leq 0$$

$$L_1 + L_2 + L_3 = L$$

Where $\mu_m$ and $\mu_n$ are the maximum values of $\{\mu_3, \mu_4, \mu_5\}$ and $\{\mu_1, \mu_2\}$ respectively.

4.4 Implementing adjusted method

Simplified optimization problem is obtained in the Section 4.3. The constraints in the optimization problem can be divided into two categories. The one is the constraint of $block_{i-1}$ (64). The other one is the constraint of $block_{i+1}$ (65).

$$\begin{cases} (v_2 - v_1)(v_2 + v_1)^2 \mu_m - L_1^2 J_m \leq 0 \\ (v_2 - v_1)(v_2 + v_1) \mu_n - L_1 A_m \leq 0 \end{cases} \tag{64}$$

$$\begin{cases}(v_2-v_3)(v_2+v_3)^2\mu_m-L_3^2J_m\leq 0\\(v_2-v_3)(v_2+v_3)\mu_n-L_3A_m\leq 0\end{cases} \tag{65}$$

When the inequality sign becomes the same sign, the boundary of the optimization problem which consist of equations (66) and (67) is obtained. Two equation about $v_2$ which are used to determine the true boundary.

$$\left(\frac{\mu_n}{A_m}\right)^2(v_2^4+v_1^4-2v_2^2v_1^2)-\frac{\mu_m}{J_m}(v_2^3+v_2^2v_1-v_2v_1^2-v_1^3)=0 \tag{66}$$

$$\left(\frac{\mu_n}{A_m}\right)^2(v_2^4+v_3^4-2v_2^2v_3^2)-\frac{\mu_m}{J_m}(v_2^3+v_2^2v_3-v_2v_3^2-v_3^3)=0 \tag{67}$$

If the left side of equation (66) is less than 0, the constraint of jerk is stricter than the acceleration. Otherwise, the acceleration constraint is stricter than jerk. Because the two equation are both quartic equations, the boundary is divided into at most 4 sections. After solving the equation, the solutions of the equation need to judge whether they are in the range $[\max(v_1,v_3),v_m]$. In each section $[a,b]$, the optimization problem become simpler. There are four situations in the section.

Situation 1:

$$\left(\frac{\mu_n}{A_m}\right)^2(v_2^4+v_1^4-2v_2^2v_1^2)-\frac{\mu_m}{J_m}(v_2^3+v_2^2v_1-v_2v_1^2-v_1^3)\leq 0 \tag{68}$$

$$\left(\frac{\mu_n}{A_m}\right)^2(v_2^4+v_3^4-2v_2^2v_3^2)-\frac{\mu_m}{J_m}(v_2^3+v_2^2v_3-v_2v_3^2-v_3^3)\leq 0$$

$$\text{Min}\quad \frac{L}{v_2}-\frac{v_1}{2v_2}\sqrt{\frac{\mu_m}{J_m}(v_2-v_1)}-\frac{v_3}{2v_2}\sqrt{\frac{\mu_m}{J_m}(v_2-v_3)} \tag{69}$$

$$a\leq v_2\leq b$$

Situation 2:

$$\left(\frac{\mu_n}{A_m}\right)^2(v_2^4+v_1^4-2v_2^2v_1^2)-\frac{\mu_m}{J_m}(v_2^3+v_2^2v_1-v_2v_1^2-v_1^3)\leq 0 \tag{70}$$

$$\left(\frac{\mu_n}{A_m}\right)^2 \left(v_2^4 + v_3^4 - 2v_2^2 v_3^2\right) - \frac{\mu_m}{J_m}\left(v_2^3 + v_2^2 v_3 - v_2 v_3^2 - v_3^3\right) \geq 0$$

$$\text{Min} \quad \frac{L}{v_2} - \frac{v_3}{v_2}\frac{\mu_n}{A_m}(v_2 - v_3) - \frac{v_1}{v_2}\sqrt{\frac{\mu_m}{J_m}(v_2 - v_1)} \tag{71}$$

$$a \leq v_2 \leq b$$

Situation 3:

$$\left(\frac{\mu_n}{A_m}\right)^2 \left(v_2^4 + v_1^4 - 2v_2^2 v_1^2\right) - \frac{\mu_m}{J_m}\left(v_2^3 + v_2^2 v_1 - v_2 v_1^2 - v_1^3\right) \geq 0 \tag{72}$$

$$\left(\frac{\mu_n}{A_m}\right)^2 \left(v_2^4 + v_3^4 - 2v_2^2 v_3^2\right) - \frac{\mu_m}{J_m}\left(v_2^3 + v_2^2 v_3 - v_2 v_3^2 - v_3^3\right) \geq 0$$

$$\text{Min} \quad \frac{L}{v_2} - \frac{v_1}{v_2}\frac{\mu_n}{A_m}(v_2 - v_1) - \frac{v_3}{v_2}\frac{\mu_n}{A_m}(v_2 - v_3) \tag{73}$$

$$a \leq v_2 \leq b$$

Situation 4:

$$\left(\frac{\mu_n}{A_m}\right)^2 \left(v_2^4 + v_1^4 - 2v_2^2 v_1^2\right) - \frac{\mu_m}{J_m}\left(v_2^3 + v_2^2 v_1 - v_2 v_1^2 - v_1^3\right) \geq 0 \tag{74}$$

$$\left(\frac{\mu_n}{A_m}\right)^2 \left(v_2^4 + v_3^4 - 2v_2^2 v_3^2\right) - \frac{\mu_m}{J_m}\left(v_2^3 + v_2^2 v_3 - v_2 v_3^2 - v_3^3\right) \leq 0$$

$$\text{Min} \quad \frac{L}{v_2} - \frac{v_1}{v_2}\frac{\mu_n}{A_m}(v_2 - v_1) - \frac{v_3}{v_2}\sqrt{\frac{\mu_m}{J_m}(v_2 - v_3)} \tag{75}$$

$$a \leq v_2 \leq b$$

There are some methods to find the optimal solution. For these solutions, finding the extreme point of the objective function is a feasible method. Because the derivative function of these four objective functions are continuous at each section, the value of the derivative function is zero at the extreme point. Then, the optimal value of $v_2$ can be obtained via comparing values of objective function at extreme points and end points of intervals.

In situation1, in order to compute the extreme point, the derivative function of objective function is equal to 0:

$$-\frac{L}{v_2^2}+\frac{v_1}{v_2^2}\sqrt{\frac{\mu_m}{J_m}}\frac{v_2-2v_1}{2\sqrt{v_2-v_1}}+\frac{v_3}{v_2^2}\sqrt{\frac{\mu_m}{J_m}}\frac{v_2-2v_3}{2\sqrt{v_2-v_3}}=0 \tag{76}$$

Then, the equation can be transformed into polynomial equation:

$$8(v_2-2v_3)^2(v_2-v_1)^2(v_2-v_3)=16L^4(v_2-v_3)^2(v_2-v_1)^2+(v_2-2v_1)^4(v_2-v_3)^2-8L^2(v_2-2v_1)^2(v_2-v_1)(v_2-v_3)^2$$
$$+(v_2-2v_3)^4(v_2-v_1)^2-2(v_2-2v_1)^2(v_2-2v_3)^2(v_2-v_1)(v_2-v_3)$$

(77)

The solution of equation (77) is also the solution of equation (76)

In situation 2, the equation to compute extreme points can be obtained:

$$-\frac{L}{v_2^2}+\frac{v_1}{v_2^2}\sqrt{\frac{\mu_m}{J_m}}\frac{v_2-2v_1}{2\sqrt{v_2-v_1}}-\frac{v_3^2}{v_2^2}\frac{\mu_n}{A_m}=0 \tag{78}$$

Similarly, the polynomial equation is obtained:

$$v_1^2\frac{\mu_m}{J_m}(v_2-2v_1)^2=\left(v_3^2\frac{\mu_n}{A_m}+L\right)^2 4(v_2-v_1) \tag{79}$$

In situation 4, the struct of objective function is the same as objective function in situation 2. Then the equation can be obtained.

$$-\frac{L}{v_2^2}+\frac{v_3}{v_2^2}\sqrt{\frac{\mu_m}{J_m}}\frac{v_2-2v_3}{2\sqrt{v_2-v_3}}-\frac{v_1^2}{v_2^2}\frac{\mu_n}{A_m}=0 \tag{80}$$

$$v_3^2\frac{\mu_m}{J_m}(v_2-2v_3)^2=\left(v_1^2\frac{\mu_n}{A_m}+L\right)^2 4(v_2-v_1) \tag{81}$$

In the situation 3, because $\mu_n\geq 0$, the value of derivative function is always less than 0.

$$-\frac{L}{v_2^2}-\frac{v_1^2}{v_2^2}\frac{\mu_n}{A_m}-\frac{v_3^2}{v_2^2}\frac{\mu_n}{A_m}\leq 0$$

It means that the objective function is decreasing with respect to the feed rate $v_2$.

Then, the process of solving optimal problems can be implemented through the following steps.

Step1: Computing the parameters $\mu_1$, $\mu_2$, $\mu_3$, $\mu_4$, $\mu_5$, and selecting the maximum parameters $\mu_n$ and $\mu_m$ from the section $\{\mu_1,\mu_2\}$ and $\{\mu_3,\mu_4,\mu_4\}$ respectively. Then, obtaining the optimization problem (63);

Step2: According to the given conditions $L_1 + L_3 \leq L(L_1 > 0, L_3 > 0)$, find the range of feed rate, remarked as $[v_l, v_h]$.

Step3: Solving the equations (66), (67) and selecting the solutions within the feed rate range. Then, dividing the feed rate range into $N\ (N \leq 8)$ sections. $[x_i, x_{i+1}], 1 < i \leq N$;

Step4: For each section, finding the optimization solutions $\{v_i\}, 1 \leq i \leq N$.

Step5: From these optimization solutions, find the global optimization solution through comparing the value of objective function at each solution $\{v_i\}, 1 \leq i \leq N$.

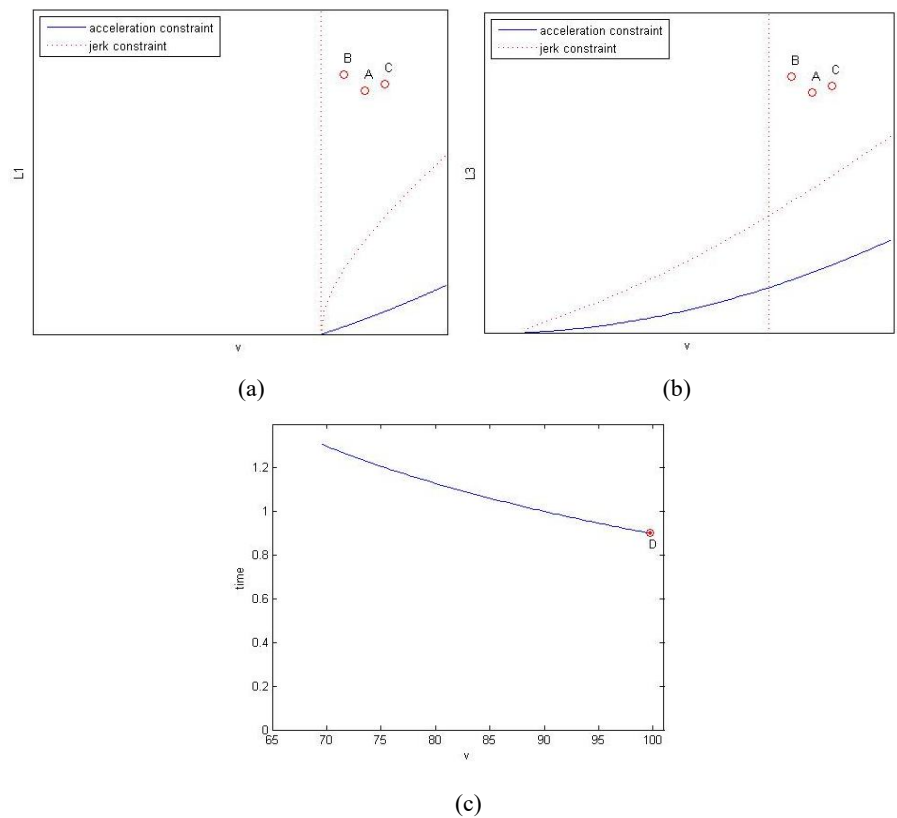

Fig. 12. (a) Feasible region for first block and points A, B and C are feasible solutions;(b) Feasible region for third block and points A,B and C are feasible solutions;(c) graph of objective function and point D is the optimal solution.

Fig. 12(a) and (b) are the schematic diagrams of constraints. The horizontal axis represents the feed rate. The vertical axis represents displacement. In Fig. 12(a) and (b), red curve and black curve represent the constraints of jerk and acceleration respectively. Points A, B and C are the feasible points. The vertical dotted line is lower boundary of feed rate. The curve in the Fig. 12(c) is objective function in the situation 1. It decreases gradually as the feed rate increases. Hence, point D is the optimal solution of this problem.

As mentioned in the Section 4.2, the proposed method is more efficient than sine feed rate scheduling method [21] in some situations. In the section 2, the limits of tangential acceleration and jerk are given as equations (14) and (15). Then, suppose that sine-curve method and proposed method are used to design the feed rate profile for the same process, $block = \{v_s, v_e, u_s, u_e, l, T, s\}$

For sine-curve profile, according to equation (12) and (13):

$$\left|\frac{v_e - v_s}{2}\right|\frac{\pi}{T} \leq A_m \tag{82}$$

$$\left|\frac{v_e - v_s}{2}\right|\left(\frac{\pi}{T}\right)^2 \leq J_m \tag{83}$$

Comparing to the equations (48), (49), (54), (56) and (58), two parameters can be computed:

$$\mu_6 = \frac{\pi}{4}, \mu_7 = \frac{\pi^2}{8} \tag{84}$$

$\mu_6$ is the parameter for acceleration constraint as the same with $\mu_n$. And $\mu_7$ is the parameter for jerk constraint just like $\mu_m$. There are 3 situations divided by the strictness of the constraints.

$\mu_n = \max\{\mu_1, \mu_2\}, \mu_m = \max\{\mu_3, \mu_4, \mu_5\}$

1. acceleration constraint is strict and jerk constraint is loose:

The shape parameter s can be assigned a real number from 2 to 2.5. If $s < 2$, $\mu_m$ will be much larger than $\mu_6$

2. acceleration constraint is loose and jerk constraint is strict:

The shape parameter s can be assigned 3.3. Then, $\mu_m = 1.1282 < \mu_7$. In other words, the proposed method can generate the feed rate profile with larger difference between start feed rate and end feed rate.

3. acceleration and jerk constraint are both strict:

In this situation, shape parameter s can also be assigned 3.3 which can ensure that the efficiency of proposed method is only slightly worse than that of s-curve method. In most actual circumstances, situation 1 and 2 are more common. In order to distinguish three situations, a displacement $\bar{l}$ can be calculated with feasible acceleration $A_m$ and S which can be take 3 using equation (36). Then, the maximum jerk $\bar{J}$ is computed using the displacement $\bar{l}$ and equation (36). Finally, if $\bar{J} < J_m$, it is situation 1. If $\bar{J} > J_m$, it is situation 2. Otherwise, it is situation 3.

For avoiding the jump and discontinuity at the junction of two blocks, it is necessary to check which situation the present blocks are. In fact, the first two situations only exist at beginning and end. Then, suppose that there are three blocks, $block_i$, $block_{i+1}$, and $block_{i+2}$, for third and fourth situations, they can be distinguish via judging whether $u_{i,e}$ of $block_i$ is equal to $u_{i+2,s}$ for $block_{i+2}$. In the fourth situation, start feed rate and end feed rate may change under special conditions. After feed rate adjusting, end feed rate $v_{i+2,e}$ of $block_{i+2}$ and the start feed rate $v_{i,s}$ of $block_i$ may be changed. If $v_{i,s}$ is changed, $v_{i-1,e}$ of feed rate $block_{i-1}$ will be replaced with $v_{i,s}$, and if $v_{i+2,e}$ is changed, $v_{i+3,s}$ will be replaced with $v_{i+2,e}$. Then, feed rate adjusting method will be implemented on the next several blocks (third situation or fourth situation). When all blocks are adjusted, if there are feed rate which are changed, above adjusting process is performed again until there are no feed rate changed.

## 5. Simulation and experimental verification

In the following, simulation and experiments are performed on the two NURBS curve to validate the proposed feed rate scheduling method. The first curve is WM-shape curve, which is a simple 2-degree NURBS curve with 8 control points. The other curve is butterfly curve, which is a complex 3-degree NURBS curve with 51 control points. For the geometric constraints, a maximum chord error is given. The kinetic constrains are maximum velocity, acceleration and jerk. Then, the proposed method is applied to generate feed rate profile. To compare the efficiency between sine-curve method and proposed method, the feed rate scheduling method according to the equations (11), (12) and (13) is applied to generate the feed rate profile for each block, after splitting the feed rate curve. The constraints of acceleration and jerk can be expressed in form of (82) and (83), and only replace parameters $\mu_n$ and $\mu_m$ with $\mu_6$ and $\mu_7$. In the interpolation stage, two-order Taylor interpolation algorithms are also performed to verify the efficiency of the proposed method with the interpolation parameters illustrated in Table 1 and Table 3. Since there is truncation error in initial two-order Taylor interpolation method, an iterative process based on dichotomy method is applied to calculate accuracy parameter u correspond to interpolation step length L. The theoretical machining time is calculated according to equation (85). Machining time is also used to compare machining efficiency.

$$T_{total} = \sum_{i=1}^{N} t_i \tag{85}$$

Where $N$ is the number of blocks, and $t_i$ is the element of $block_i$ after feed rate adjusting.

The simulations are conducted on a personal computer with Intel(R) Core (TM) i7-6500U 2.59-GHz CPU, 8.00-GB SDRAM, and Windows 10 operating system. All the algorithms for the simulations are developed and implemented on Dev-C++ by C language.

Table 1 Parameters of kinematic constraints and feed rate scheduling method for WM curve

| Parameters | Symbols | Units |
|---|---|---|
| Sampling time | $T_s$ | 1ms |
| Chord error | $\delta$ | 0.0005mm |
| Maximum feed rate | $V_{max}$ | 100mm/s |
| Maximum acceleration | $A_{max}$ | 1000mm/s² |
| Maximum jerk | $J_m$ | 26000mm/s³ |
| Shape parameter | $S$ | 3.3 |

5.1 Simulation results of WM-shape curve

The first NURBS curve is a WM-shape curve as shown in Fig.. 13(a) and the parameters of simulations are listed in Table 1. The feed rate curve under the chord error is shown in the Fig. 13(b) which illustrates that the proposed algorithm how to detect the breaking points. The red points are breaking points which also contain start and end points of NURBS curve. Totally, there are 20 breaking points and 19 blocks. Between two adjacent points, the process is one of three types acceleration, deceleration and constant feed rate. As the given parameters, it is classified as situation 2. Then, the method mentioned in Section 4 is used to find optimal solution of optimal problem (43) or (46), for each block. Finally, the feed rate profile generated by proposed method is planned as shown in Fig. 13(c). As shown in Fig. 13(d)-(f), the profiles of acceleration, jerk and chord error generated by the proposed method are almost constrained on the values of $1000 mm/s^2$, $26000 mm/s^3$ and 0.5 μm, respectively. From the Fig. 13 (i), sine-curve method exceeds jerk limitation a little and proposed method strictly complies with jerk constraint.

In the Fig. 13(g)-(j), the profiles of feed rate, acceleration and jerk generated by sine-curve method are planned. Comparing proposed method with sine-curve method, the feed rate value at point A is 100mm/s in the Fig. 13(c) but the feed rate at same point is 97mm/s in the Fig. 13(g). And in the first block and the end block which are constant feed rate with maximum feed rate, the displacement l for first block and end block are 7.38 and 9.96mm, respectively using the proposed method. However, two displacement of first block and end block are 7.14 and 8.22mm, respectively using sine-curve method. For maximum acceleration, it can reach 928mm/$s^2$ in proposed method and is just 768mm/$s^2$ in sine-curve method. Finally, after interpolation calculation, there are 701 interpolation points in the results of proposed method and 722 interpolation points in the results of sine-curve method.

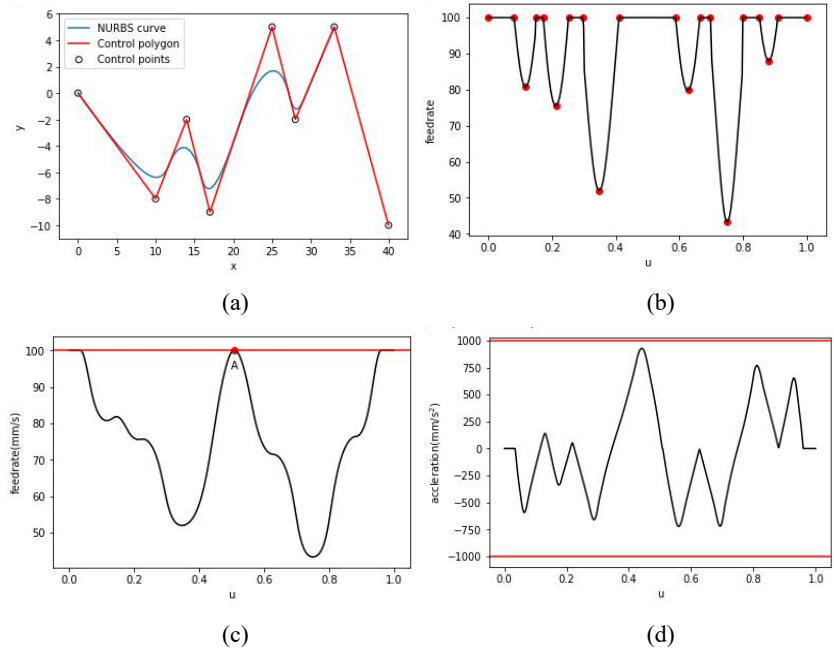

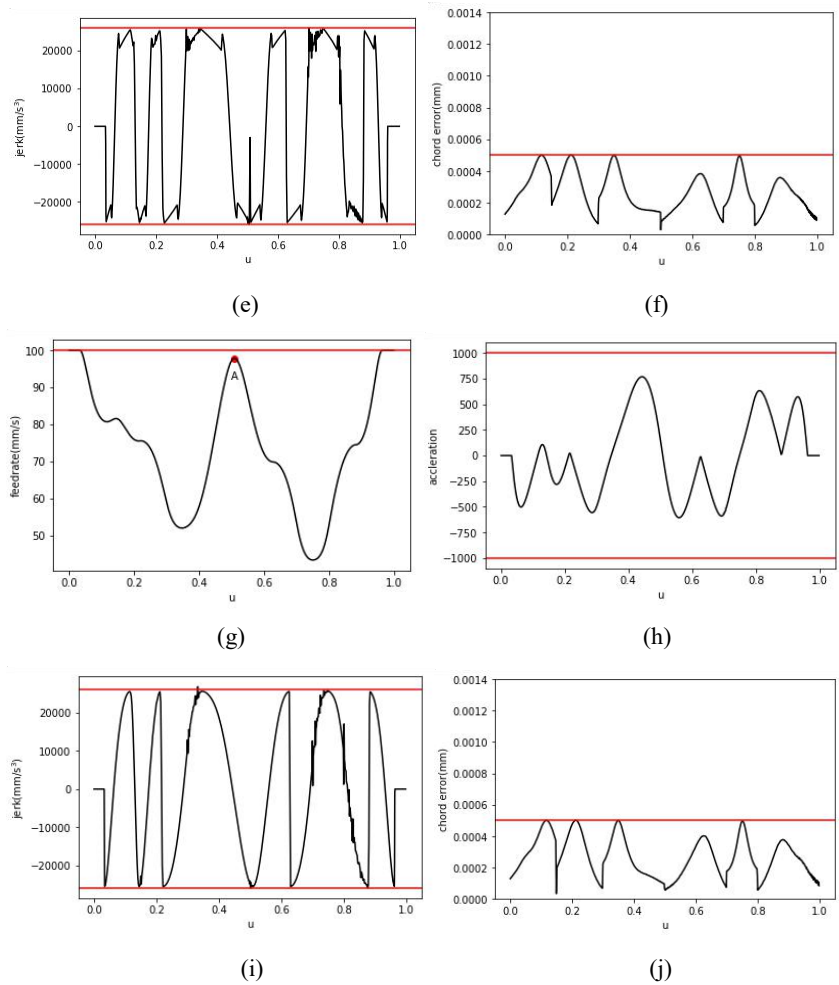

(e) (f)

(g) (h)

(i) (j)

Fig. 13.Simulation results for WM-shape curve.(a)Feed rate profile confined on chord error and breaking points,(b)WM-shape curve,(c)Feed rate profile of proposed method,(d)Acceleration profile of proposed method,(e)Jerk profile of proposed method,(f)Chord error profile of proposed method,(g)Feed rate profile of sine-curve method,(h)Acceleration profile of sine-curve method,(i)Jerk profile of sine-curve method,(j)Chord error profile of sine-curve method.

The simulation results show that not only proposed method can confine the chord error under the 0.5 μm but also the acceleration and jerk are bounded. Besides, in high-speed machining, efficiency of proposed method is slightly improved compared to sine-curve method.

Table 2: Simulation results of WM-shape curve

| Method | Max feed rate | Max acceleration | Max jerk | Max chord error | Total Time | Number of points |
|---|---|---|---|---|---|---|
| Proposed | 100mm/s | 928mm/s$^2$ | 25986mm/s$^3$ | $5 \times 10^{-4}$mm | 0.702s | 701 |
| Sine-curve | 100mm/s | 768mm/s$^2$ | 26772mm/s$^3$ | $5 \times 10^{-4}$mm | 0.721s | 722 |

Table 3: Parameters of kinematic constraints and feed rate scheduling method for butterfly curve.

| Parameters | Symbols | Values |
|---|---|---|
| Sampling time | $T_s$ | 1ms |
| Chord error | $\delta$ | 0.0005mm |
| Maximum feed rate | $V_{max}$ | 100mm/s |
| Maximum acceleration | $A_{max}$ | 3000mm/s$^2$ |
| Maximum jerk | $J_{max}$ | 55000mm/s$^3$ |
| Shape parameter | $S$ | 3.3 |

## 5.2 Simulation results of butterfly curve

The second NURBS curve is a butterfly curve as shown in Fig. 14(b) and the parameters of simulations are listed in Table 3. In the stage of curve scanning, the breaking points, which are red points in the Fig. 14(a), are determined. Through the feed rate curve scanning, 37 breaking points are determined using the strategy in Section 3. Then, totally 36 blocks are divided via these breaking points. According to given parameters, $\mu_m$ and $\mu_n$ can be calculated. Then, the solutions of equation (43) or (46) is computed. According to the given parameters, jerk constraints is strict. Thus, the time-optimal feed rate adjusting can be implemented for each block, using the method mentioned in the Section 4. After feed rate adjusting, some blocks will degrade into points which the start parameters are equal to end parameters. For example, the points where the parameter u are 0.346605 and 0.653272 respectively are degenerated from the original blocks, in the Fig. 14(c) and (g).

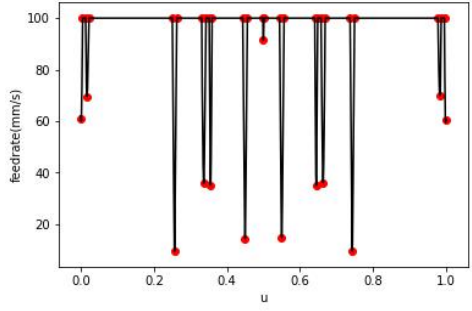
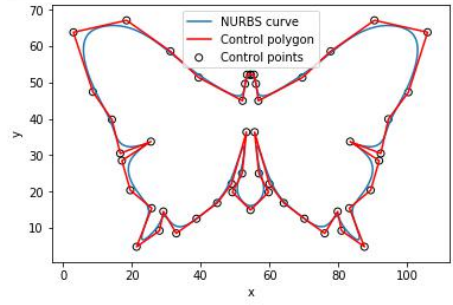

(a)　　　　　　　　　　　　　(b)

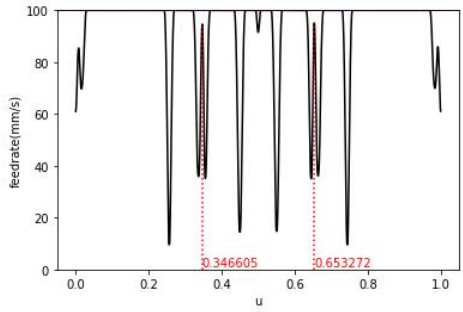
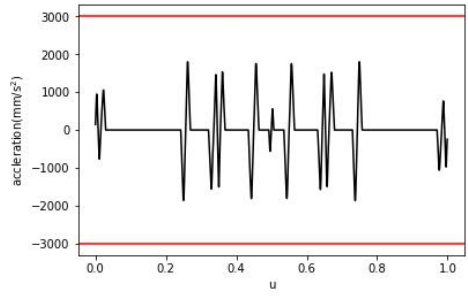

(c)　　　　　　　　　　　　　(d)

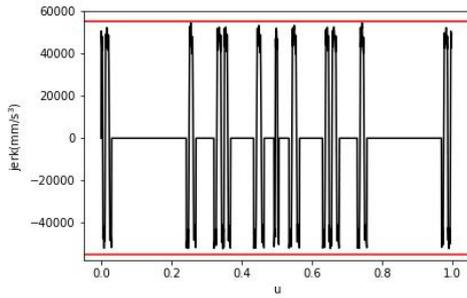
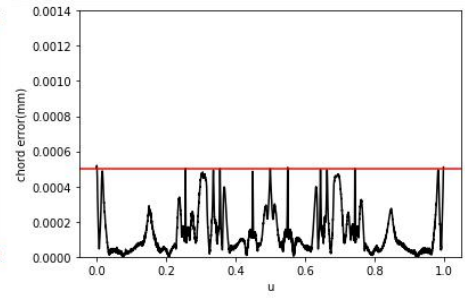

(e)　　　　　　　　　　　　　(f)

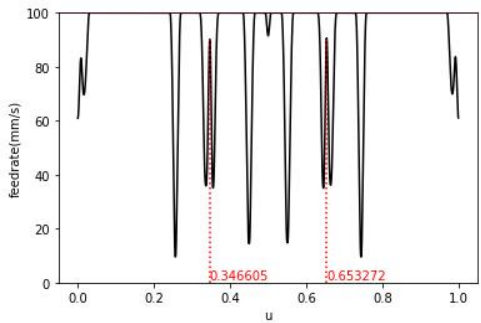
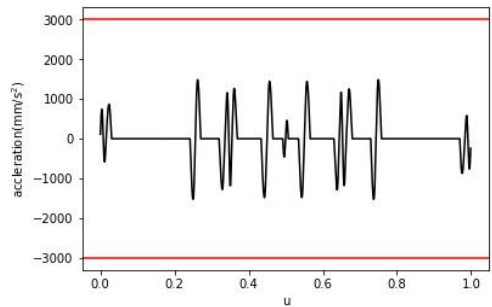

(g)　　　　　　　　　　　　　(h)

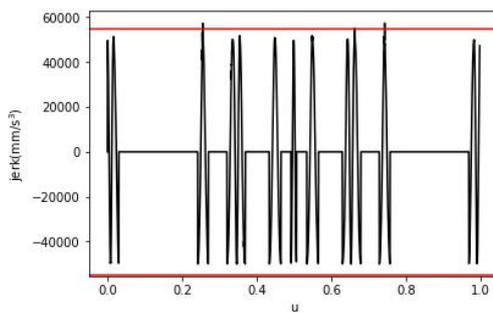
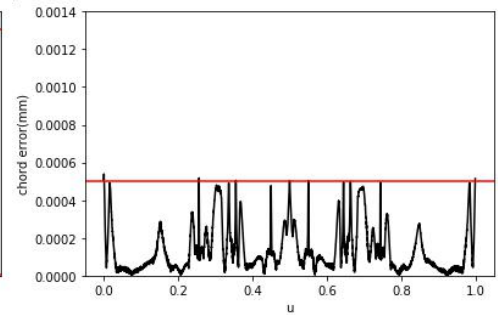

(i)　　　　　　　　　　　　(j)

Fig. 14(a)Feed rate profile confined on the chord error and breaking points,(b)Butterfly curve,(c)Feed rate profile of proposed method,(d)Acceleration profile of proposed method,(e)Jerk profile of proposed file,(f)Chord error profile of proposed method,(g)Feed rate profile of sine-curve method,(h)Acceleration profile of sine-curve method,(i)Jerk profile of sine-curve method,(j)Chord error of sine-curve method.

In the Fig. 14(d)-(f), profile of acceleration, jerk and chord error generated by proposed method are shown and almost constrained on the values of $3000mm/s^2$, $55000mm/s^3$ and 0.5 μm. Meanwhile, the feed rate profile generated by proposed method is shown in Fig. 14(c). The Fig. 14(g)-(j) show the profile of feed rate, chord error, acceleration and jerk generated by sine-curve method. Comparing Fig. 14(c) and Fig. 14(g), the total time of blocks in Fig. 14(c) and Fig. 14(g) are 4.360 and 4.417s, respectively. At the points where parameter u are 0.346605 and 0.653272 respectively, the feed rate at these points are 95.806648 and 96.284721 mm/s. The feed rate generated by sine-curve method at these points are 93.044418 and 93.507507 mm/s. In the Fig. 13 (e) and (i), maximum value of jerk in sine-curve method is greater than jerk limitation, and proposed method can strictly meet jerk constraint. After interpolation calculation, there are 4360 points using the proposed feed rate scheduling method and there are 4420 points using sine-curve method. Then, the simulation results for butterfly curve are presented in the Table 4.

Table 4 Simulation results of sine-curve method

| Method | Max feed rate | Max acceleration | Max jerk | Max chord error | Total Time | Number of points |
| --- | --- | --- | --- | --- | --- | --- |
| **Proposed** | 100mm/s | $1833mm/s^2$ | $54989mm/s^3$ | $5.19\times10^{-4}$ mm | 4.360s | 4360 |
| **Sine-curve** | 100mm/s | $1554mm/s^2$ | $63546mm/s^3$ | $5.39\times10^{-4}$ mm | 4.417s | 4420 |

5.3 Discussion

The chord error profiles of two NURBS curve are confined on the chord error tolerance. From the Fig. 13(f), (j)and Fig. 14(f), (j), it can be seen that feed rate determining method which is proposed in Section3 is able to determine the feed rate satisfying chord error constraints. Then, for feed rate profile demonstrated in Fig. 13(a) and Fig. 14(a), breaking points for two NURBS curves are determined with suitable parameter $\mu_s$. In application, the value of $\mu_s$ should be consistent with NURBS curve. Using (21), the boundary of screening factor are obtained, and parameter $\mu_s$ should be slightly greater lower boundary to ensure all breaking points are included. From the table 2 and table 4, the kinematic characteristics containing acceleration and jerk are confined on the given value. Therefore, the proposed feed rate scheduling method based on Sigmoid function can guarantee that kinematic characteristics does not exceed the preset value. Meanwhile, sine-curve method is also used to design feed rate changing process for each block. The maximum values of jerk are 26772mm/$s^3$ and 63546mm/$s^3$, respectively for sine-curve method. In repeated simulations with different jerk value, sine-curve method has unstable performance, and there are always different points exceeding the jerk limitation with the deceleration of jerk limitation. It may be caused by two-order Taylor method which is not very accurate because of truncation error. Thus, a more accurate interpolation method is necessary for sine-curve method. Comparing two method, the theoretical machining time of proposed method is shorter than sine-curve method for two NURBS curve. And the number of interpolation points for proposed method is less than sine-curve method. According to the data, proposed feed rate scheduling method has a certain extent of advantages in the term of machining efficiency compared with sine-curve method. Besides, proposed method can reach higher acceleration than sine-curve method. It indicates that kinematic constraints can be more fully used in proposed method than sine-curve method.

**6 Conclusion**

This paper proposes a Sigmoid function-based feed rate scheduling method with chord error constraints and kinematic constraints. An approximate relationship between feed rate and chord error is used to calculate accurate feed rate value in pre-interpolation process. Through the pre-interpolation process, the shape of feed rate curve with chord error constraint is determined. Then, a two-step scanning algorithm, which aims to find breaking point, is carried out to scanning whole feed rate curve, and feed rate curve will be divided into three kinds of blocks: acceleration, deceleration and constant feed rate block according the breaking point. In every block, the feed rate profile will be designed by Sigmoid function with compounding two linear functions and local polynomial fitting. The feed rate profile is "one master, two slaves" method of which the connection of three part is closer than polynomial method, and expression ability is better than trigonometric method. Therefore, proposed method is more convenient to be applied than polynomial method and has a certain of advantage in efficiency than sine-curve method. Since kinematic constraints are also needed to be considered, a time-optimal formulate is established to adjust the feed rate value at breaking points. Although the proposed method is jerk-limited method, it can become jerk-continues by using high-order polynomial to fit. Meanwhile, contour error and kinematic characteristics of each axis are need considered. Both two aspects will be introduced in future work.

# 7 Acknowledgement

This work has been supported by the National Key Research and Development Program of China (Grant No.2018YFB1107402) and the National Natural Science Foundation of China (Grants No. 12001028 and No.11290141).